\documentclass[12pt,onecolumn]{revtex4}

\usepackage{amssymb}
\usepackage{amsmath}
\usepackage{graphics}
\usepackage{latexsym}
\usepackage{array}
\usepackage[activeacute,english]{babel}
\usepackage[latin1]{inputenc}
\usepackage[dvips]{graphicx}
\usepackage{color}

\newcommand{\bn}{\begin{eqnarray}}
\newcommand{\en}{\end{eqnarray}}
\newcommand{\bq}{\begin{equation}}
\newcommand{\eq}{\end{equation}}

\newcommand{\bc}{\begin{center}}
\newcommand{\ec}{\end{center}}

\begin{document}
\title{\bf Some classical properties of the non-abelian Yang-Mills theories}
\author{J. A. S\'anchez-Monroy$^{1}$}
\author{C. J. Quimbay$^{1,2}$}
\affiliation{$^{1}$Grupo de Campos y Part\'\i culas, Universidad Nacional de Colombia, Sede Bogot\'a\\
$^{2}$Associate researcher of CIF, Bogotá, Colombia.}

\date{\today}

\begin{abstract}
We present some classical properties for non-abelian Yang-Mills
theories that we extract directly from the Maxwell's equations of
the theory. We write the equations of motion for the $SU(3)$
Yang-Mills theory using the language of Maxwell's equations in
both differential and integral forms. We show that vectorial gauge
fields in this theory are non-fermionic sources for non-abelian
electric and magnetic fields. These vectorial gauge fields are
also responsible for the existence of magnetic monopoles. We build
the continuity equation and the energy-momentum tensor for the
non-abelian case.
\\[0.02in]
\textit{Keywords}: Yang-Mills theory; Maxwell's equations; integral and differential forms; magnetic monopoles.
\begin{center}
\textbf{Resumen}
\end{center}

En este artículo se presentan algunas propiedades clásicas de las teorías
de Yang-Mills no abelianas, que se extraen directamente de las ecuaciones
de Maxwell de la teoría. Obtenemos las ecuaciones de movimiento para una
teoría de Yang-Mills del grupo $SU(3)$ en su forma diferencial e integral,
utilizando el lenguaje de las ecuaciones de Maxwell. Mostramos que los campos
gauge en esta teoría son fuentes no fermiónicas para campos eléctricos y magnéticos
no abelianos. Estos campos de gauge son responsables de la  existencia de monopolos
magnéticos. Finalmente, se construyen la ecuación de continuidad y el tensor
energía-impulso para el caso no abeliano.
\\[0.02in]
\textit{Descriptores}: Teorías de Yang-Mills; ecuaciones de Maxwell; forma diferencia e integral; monopolos magnéticos.
\\[0.02in]
PACS:03.50.-z; 03.50.Kk
\end{abstract}

\maketitle

\section{Introduction}

In the context of relativistic quantum theory of electromagnetism,
the interaction among two electrically charged particles is
mediated through the exchange of virtual photons. Photons are the
quantum excitations of electromagnetic field, which is a vectorial
gauge field invariant under $U(1)$ abelian transformations.
Similarly, the strong and weak interactions are described by means
of non-abelian vectorial gauge fields. Field theories describing
the behavior of pure vectorial gauge fields are known as
Yang-Mills theories. Symmetries and properties of Yang-Mills
theories are basic ingredients for the theoretical treatment of
the fundamental interactions between elementary particles. The
study of the classical properties of the Relativistic
Electrodynamics, a particular case of abelian Yang-Mills theory,
is a well known topic in literature \cite{Ja9,ed}. On the other
hand, the classical properties of non-abelian Yang-Mills theories
is a subject less studied in the field theory literature. But it
is possible to find some books which study this subject
\cite{covari,valery,boris,Theodore}. These books give a similar
emphasis to the presentation of Yang-Mills equations as functions
of gauge potentials, however the introduction of these equations
in terms of non-abelian electric and magnetic fields is
practically absent \cite{Theodore,Jackiw,Jackiw1,Faddeev}.
\par
The main goal of this paper is to present some classical
properties for non-abelian Yang-Mills theories. We can extract
these properties writing the equation of motion for non-abelian
Yang-Mills theories, using the language of electric and magnetic
fields. We write these non-abelian Maxwell's equations in both
differential and integral forms as it is usual for Maxwell's
equations of Classical Electrodynamics. We restrict our interest
to the case of the $SU(3)$ Yang-Mills theory, however the analysis
is the same for any group $SU(N)$. We show that non-abelian
electric and magnetic fields can be generated by vectorial gauge
fields. These vectorial gauge fields are also responsible for the
existence of magnetic monopoles. Finally, we build the continuity
equation and the energy-momentum tensor for the non-abelian case.

\section{Non-abelian Maxwell's equations}
Non-abelian gauge theories have some differences respect to the
abelian ones, as for instance the existence of a multiplicity of
gauge fields, self-interactions, and gauge transformations that
involve the gauge fields \cite{trans}. Particularly these
differences are clearly observed if we contrast the $SU(3)$
Yang-Mills theory including color charge sources with the
Relativistic Classical Electrodynamics including electric charge
sources. The $SU(3)$ Yang-Mills theory is described by the
following Lagrangian density:
\begin{equation}\label{lagran}
\mathcal{L}=-\frac1 {4}F^a_{\mu \rho }F_a^{\mu \rho }+ gJ^{\mu}_a
A^a_{\mu},
\end{equation}
where
\begin{eqnarray}\label{ym1}
F^a_{\mu \nu}&=&\partial _\mu A^a _\nu -\partial _\nu A^a_\mu +
gC^a_{bc}A^b_\mu A^c_\nu,
\end{eqnarray}
is the non-abelian field strength tensor, $A^a_\mu$ the gluon
fields, $J^\mu_a$ the color charge sources, $C^a _{bc}$ the
structure constants of Lie algebra associated to the $SU(3)$ gauge
group, $g$ the running coupling constant and $a, b, c=1, 2 , ...,
8$. The first term of Eq. (\ref{lagran}) describes the kinetic
energy of eight gluon fields and their respective
auto-interactions. The second term describes the interactions of
the gluon fields with the fermionic fields. Applying the
variational method of the Classical Field Theory, we obtain that
the equations of motion for the SU(3) Yang-Mills theory with color
charge sources are
\begin{equation}\label{yms}
\partial _\mu F_b ^{\mu
\nu}+gC^a _{bc}A^c _\mu F_a ^{\mu \nu}=gJ_b^\nu=g\bar \psi \gamma
^\nu \lambda_b\psi,
\end{equation}
where $\lambda_b$ are the Gell-Mann matrices and $\psi$ the
fermionic fields. The equations of motion given by Eq. (\ref{yms})
represent a system of non-lineal equations.
\par
The electric ($\vec E$) and magnetic ($\vec B$) fields, in the
Classical Electrodynamics, are defined as the components of the
electromagnetic field strength tensor ($F_{\mu \nu}$), in the
following form:
\begin{eqnarray}
E_i&:=&F_{0 i}=-F^{0 i}=-F_{i 0},\\
B_n&:=&-\frac{1}{2}\varepsilon _{nij}F^{ij},
\end{eqnarray}
being $n,i,j=1,2,3$. Starting from the Lagrangian density of the
electromagnetic field with sources it is possible to obtain the
Yang-Mills equations using the Euler-Lagrange equations. From
these equations it is possible to obtain the homogeneous Maxwell's
equations.

In an analogous way, we consider the non-abelian Maxwell's
equations for the SU(3) Yang-Mills theory with color charge
sources. The non-abelian field strength tensor is given by Eq.
(\ref{ym1}), where the covariant non-abelian gauge field is
written as $A^a_\mu=(A^{0 a}, -\vec A^{a})$. The electric and
magnetic color fields are defined respectively as
\begin{eqnarray}
E^a_i&:=&F^a_{i 0}=\partial _i A^a _0 -\partial _0 A^a_i +
gC^a_{bc}A^b_i A^c_0,\label{ecf}\\
B^{aj}&:=&-\frac{1}{2}\varepsilon ^ {jik}F^a_{ik}=-\frac{1}{2}
\varepsilon ^{jik}\left(\partial _i A^a _k -\partial _k A^a_i
+gC^a_{bc}A^b_i A^c_k\right)\label{mcf},
\end{eqnarray}
being $n,i,j=1,2,3$. In vectorial notation, the electric and
magnetic color fields are \cite{Jackiw}
\begin{eqnarray}
\vec{E}^a&=&-\vec{\nabla} A_0^a-\partial_t \vec{A}^a+gC^a_{bc}
A_0^b\vec{A}^c,\\
\vec{B}^a&=&\vec{\nabla} \times \vec{A}^a-\frac{1}{2}gC^a_{bc}
\left(\vec{A}^b \times \vec{A}^c\right).
\end{eqnarray}
In contrast with the magnetic field of Electrodynamics, the
magnetic color field for SU(3) Yang-Mills theory can be written as
the sum of a rotor term and a non-rotor term.

Using the definition Eq. (\ref{ecf}) in the Eq. (\ref{yms}), we
obtain that the first Maxwell's equation for the SU(3) Yang-Mills
theory with color charge sources is given by \cite{Theodore,Jackiw,Jackiw1,Faddeev}
\begin{equation}
\partial_i E^a_{i}+gC^a_{bc}A_{bi} E^c_{i}=g\rho^a,
\end{equation}
or in vectorial notation
\begin{equation}\label{me1}
\vec{\nabla} \cdot \vec{E}^a=gC^a_{bc}\vec{A}^b \cdot
\vec{E}^c+g\rho^a,
\end{equation}
where we have used the fact that the color charge source can be
written as $J^a_\mu=(\rho^a, -\vec J^a)$ and $A^a_\mu=(A^{0 a},
-\vec A^{a})$. Reasons by the magnetic color field can be written
as
\begin{eqnarray}
B^{aj}&=&-\frac{1}{2}\varepsilon ^{jik}F^a_{ik},\notag\\
\varepsilon _{jpq}B^{aj}&=&-\frac{1}{2}\varepsilon _{jpq}
\varepsilon ^{jik}F^a_{ik},\notag\\
\varepsilon _{jpq}B^{aj}&=&-\frac{1}{2}(\delta_{p}^i
\delta_{q}^k-\delta_{p}^k\delta_{q}^i)F^a_{ik},\notag\\
-\varepsilon _{jpq}B^{aj}&=&F^a_{pq}=F^{apq},
\end{eqnarray}
then, the second Maxwell's equation is given by
\begin{eqnarray}
\partial^\mu F^a_{\mu j}&=&-gC^a_{bc}A^{b \mu} F^c_{\mu j}+gJ^a_j,\notag\\
\partial^0 E^a_{j}+\partial^i F^a_{i j}&=&-gC^a_{bc}A^{b0} E^c_{j}
-gC^a_{bc}A^{bi} F^c_{i j}+gJ^a_j,\notag\\
\partial^0 E^a_{j}-\partial^i \varepsilon _{lij}B^{al}&=&-gC^a_{bc}A^{b0}
E^c_{j}+gC^a_{bc}A^{bi} \varepsilon _{lij}B^{cl}+gJ^a_j,\\
\partial^0 E^a_{j}+\varepsilon _{jil}\partial^i B^{al}&=&-gC^a_{bc}A^{b0}
E^c_{j}+gC^a_{bc}A^{bi} \varepsilon _{jil}B^{cl}+gJ^a_j,
\end{eqnarray}
which in vectorial notation can be written as \cite{Theodore}
\begin{equation}\label{me2}
\vec{\nabla} \times \vec{B}^a-\partial_t\vec{E}^a=g\vec{J}^a+gC^a_{bc}A^b_0
\vec{E}^c-gC^a_{bc}\vec{A}^b\times \vec{B}^c.
\end{equation}
The other two Maxwell's equations are obtained from the
definitions of the electric color fields, given by Eq.
(\ref{ecf}), and the magnetic color field, given by Eq.
(\ref{mcf}). These Maxwell's equations are \cite{Theodore}:
 \begin{eqnarray}\label{me3}
\vec{\nabla} \cdot \vec{B}^a&=&-\frac{1}{2}gC^a_{bc}\nabla \cdot
{(\vec{A}^b \times \vec{A}^c)}
\end{eqnarray}
and
 \begin{eqnarray}\label{me4}
\vec{\nabla} \times
\vec{E}^a+\partial_t\vec{B}^a&=&-\frac{1}{2}gC^a_{bc}
\partial_t\left(\vec{A}^b \times \vec{A}^c\right)+gC^a_{bc}
\left[\vec{\nabla} \times (A^b_0 \vec{A}^c)\right].
\end{eqnarray}

We observe that Maxwell's equations for the SU(3) Yang-Mills
theory with color charge sources, which are given by Eqs.
(\ref{me1}), (\ref{me2}), (\ref{me3}) and (\ref{me4}), do not only
depend on $\vec{E}^a$ and $\vec{B}^a$ but as well on $\vec{A}^a$
and $A^a_0$. It is clear that for the abelian Yang-Mills theory
case, these equations do not have the dependence observed before.
If we put $J^a_\mu=(\rho^a, -\vec J^a)=0$ in the Eqs. (\ref{me1})
and (\ref{me2}), we observe the presence of sources of electric
and magnetic color fields whose origin are the gluon fields. It is
possible to see as the bosonic fields are charged and
simultaneously are sources of magnetic fields, i. e. the gluonic
fields have color charge. Additionally, as the divergence of
$\vec{B}^a$ non vanishing then there exist color magnetic
monopoles and the sources are the gluons but not the quarks. It is
also possible to see that the electric and magnetic color fields
are not gauge invariant and therefore they have not physical
meaning.

So as the classical electrodynamics predicts the existence of
electromagnetic waves, the SU(3) Yang-Mills theory predicts the
existence of non-abelian waves associated to strong interaction.
These waves are the solutions of the following wave equations that
can be obtained from the Maxwell's equations (see Appendix):
\begin{eqnarray}
\Delta A^a _\nu&=&gC^a _{bc}A^b _\mu (\partial _\nu A^c_\mu-
2\partial _\mu A^c _\nu-gC^c_{mn}A^m_\mu A^n_\nu).
\end{eqnarray}

The obtained Maxwell's equations can be extended directly for any
non-abelian Yang-Mills theory, taking into account that for a
$SU(N)$ gauge group there are $N^2-1$ generators, being $N$ the
group dimension. By this reason, the Maxwell's equations for a
non-abelian Yang-Mills theory are given by the Eqs. (\ref{me1}),
(\ref{me2}), (\ref{me3}) and (\ref{me4}), taking $a, b, c = 1, 2,
..., (N^2 - 1)$. For the case in which $\vec{E}^a$ and $\vec{A}^a$
are independent fields and if there exist particular boundary
conditions in the problem, the solutions of the Maxwell's
equations for a non-abelian Yang-Mills theory are unique
\cite{mate}.

In a similar way as in the abelian Yang-Mills theory case, it is
required that the non-abelian gauge fields are transformed as
\begin{eqnarray}
A'_{\mu}(x)=U(x)(A_{\mu}-ig^{-1}\partial_{\mu})U^{-1}(x),
\end{eqnarray}
where $U(x)=e^{-ig\lambda_a\chi^a(x)}$. The transformation of the
non-abelian field strength tensor has the form
\begin{eqnarray}
F'_{\mu \nu}(x)=U(x)F_{\mu \nu}U^{-1}(x).
\end{eqnarray}
If we now consider an infinitesimal gauge transformation given by
\begin{eqnarray}
U(x)\approx I-ig\lambda_a\chi^a(x),
\end{eqnarray}
it is easy to prove that the non-abelian field strength tensor is
transformed as
\begin{eqnarray}\label{infi}
F'^a_{\mu \nu}(x)=F^a_{\mu \nu}+gC^a_{b c}\chi^b F^c_{\mu \nu}.
\end{eqnarray}
This tensor is only invariant for the abelian case. Starting from
Eq. (\ref{infi}), it is possible to find that the electric and
magnetic color field can be written as
\begin{eqnarray}
E'_i&:=&F'_{i0}=E^a_{i}+gC^a_{b c}\chi^b E^c_{i},\\
B'_n&:=&-\frac{1}{2}\varepsilon _{nij}F'^{ij}=B^a_{n}+gC^a_{b c}\chi^b B^c_{n},
\end{eqnarray}
obviously the electric and magnetic color fields are not gauge
invariant. For this reason, these field do not have any physical
meaning. In a similar way as scalar and vectorial potentials are
auxiliary constructions in the Electrodynamics, the electric and
magnetic color fields do not represent measurable quantities in
the non-abelian Yang-Mills theory. For these theories, it is
possible to identify two scalar invariant quantities, which can be
written down using the physical fields in $(3 + 1)$ dimensions.
These quantities are
\begin{eqnarray}
F^a_{\mu \nu}F^{a \mu \nu}&=&2(B^2-E^2),\\
\epsilon^{\mu \nu \alpha \beta}F^a_{\mu \nu}F^a_{\alpha
\beta}&=&\vec{B}\cdot\vec{E}.
\end{eqnarray}

Below we present the integral form of non-abelian Maxwell's
equations. We first integrate the equations (\ref{me1}) and
(\ref{me3}) over the volume of a three-dimensional domain $V$
enclosed by a surface $\partial V$, and we apply the
Gauss-Ostrogradski theorem. We obtain that the integral form for
the equations (\ref{me1}) and (\ref{me3}) is given by
\cite{Jackiw,Jackiw1}

\begin{equation}\label{me1I}
\oint_{\partial V} \vec{E}^a\cdot d\vec{S} =gC^a_{bc}\int_V \vec{A}^b
\cdot \vec{E}^c dV + \int_V g\rho^a dV,
\end{equation}
and
 \begin{eqnarray}\label{me3I}
\oint_{\partial V} \vec{B}^a \cdot d\vec{S} &=&-\frac{1}{2}gC^a_{bc}
\oint_{\partial V} {(\vec{A}^b \times \vec{A}^c)} \cdot d\vec{S}.
\end{eqnarray}

We consider a two-dimensional surface $S$ which is bounded by a
loop $L$. We calculate the flux of the vector equations
(\ref{me2}) and (\ref{me4}) through $S$ and apply the Stokes
theorem. We obtain that the integral form for the equations
(\ref{me2}) and (\ref{me4}) is given by

\begin{equation}\label{me2I}
\oint_{L} \vec{B}^a \cdot d\vec{l} -\frac{d}{dt}\int_{S} \vec{E}^a
\cdot d\vec{S} =g\int_{S} \vec{J}^a \cdot d\vec{S}
+gC^a_{bc} \int_{S} A^b_0 \vec{E}^c \cdot d\vec{S}
-gC^a_{bc} \int_{S} \vec{A}^b\times \vec{B}^c \cdot d\vec{S},
\end{equation}
and
 \begin{eqnarray}\label{me4I}
\oint_{L} \vec{E}^a \cdot d\vec{l} +\frac{d}{dt} \int_{S} \vec{B}^a
\cdot d\vec{S} &=&
-\frac{1}{2}gC^a_{bc} \frac{d}{dt} \int_{S} \left(\vec{A}^b \times
\vec{A}^c\right) \cdot d\vec{S} +gC^a_{bc}
\left[\oint_{L} (A^b_0 \vec{A}^c) \cdot d\vec{l} \right].
\end{eqnarray}
The equations (\ref{me1I}), (\ref{me3I}), (\ref{me2I}) and
(\ref{me4I}) represent the integral form of non-abelian Maxwell's
equations.
\par
One of the most remarkable results in theoretical physics is
provided by Noether's theorem which establishes a relationship
among symmetries of a given action and conserved quantities of the
system described by the action \cite{noe}. This theorem is very
important for classifying the general physical characteristics of
quantum field theories \cite{trans}. In the electromagnetism there
are only positive and negative electrical charges labeled
different kinds of matter that respond to the electromagnetic
field. Additional sorts of charge are required to label particles
which respond to nuclear forces. With a variety of charges as
presented in the strong interaction, there are many more
possibilities for conservation than the one obtained from the sum
of all positive and negative electric charges \cite{covari}. This
fact can be examined by the continuity equation which is obtained
from the non-abelian Maxwell's equations. Taking the divergence of
the Eq. (\ref{me2}) and the time derivation of the Eq.
(\ref{me1}), we can obtain the following expression
\begin{eqnarray}
\partial_t\rho^a+\vec{\nabla} \cdot \vec{J}^a=-C^a_{bc}
[\partial_t(\vec{A}^b \cdot \vec{E}^c)-\vec{\nabla} \cdot (A^b_0
\vec{E}^c)+\vec{\nabla} \cdot (\vec{A}^b\times \vec{B}^c)].
\end{eqnarray}
Finally we can write the energy-momentum tensor for the Yang-Mills
fields as \cite{covari,boris,RMF,Fadbook}
\begin{eqnarray}
\Theta_{\mu \nu}=\frac{1}{4}\left(F_{a \mu}^{\
\alpha}F_{a \alpha \nu}+ \frac{\eta_{\mu \nu }}{4}F_{\alpha
\beta}^aF_a^{\alpha \beta }\right).
\end{eqnarray}
Let us to give the physical meaning of the various components of
$\Theta_{\mu \nu}$. The component $\Theta_{00}$, given by
\cite{Jackiw,Jackiw1,Faddeev}
\begin{eqnarray}
\Theta_{00}=\frac{1}{8}\left(\vec{E}^a\cdot \vec{E}_a +
\vec{B}^a\cdot \vec{B}_a\right),
\end{eqnarray}
is interpreted as the energy density. The component $\Theta_{i0}$,
given by
\begin{eqnarray}
\Theta_{i0}=\frac{1}{4}\left(\vec{E}^a\times \vec{B}_a \right),
\end{eqnarray}
represents the momentum density for the non-abelian field. Finally
the component $\Theta_{ii}$, given by
\begin{eqnarray}
\Theta_{ij}=\frac{1}{4}\left(E_i^a E_{aj}+B_i^a B_{aj}
+\frac{\delta_{ij}}{2}\left(\vec{E}^a\cdot \vec{E}_a +
\vec{B}^a\cdot \vec{B}_a\right)  \right),
\end{eqnarray}
represents the $j$th component of the momentum flow in the
non-abelian field through a unit surface perpendicular to the
$x_i$-axis. We observe that the energy-momentum tensor in the
non-abelian case has the same interpretation as in electrodynamics
\cite{Ja9,covari,boris}.


\section{Conclusions}

In this paper we have obtained the equations of motion for the
SU(3) Yang-Mills theory including color charge sources in an
analogue way as the Maxwell's equations are obtained for the
electrodynamics including electric charge sources. These
non-abelian Maxwell's equations have been obtained for the SU(3)
Yang-Mills theory, but they are directly extended for a SU(N)
Yang-Mills theory. We have found that Maxwell's equations do not
only depend on $\vec{E}^a$ and $\vec{B}^a$ but also on $\vec{A}^a$
and $A^a_0$. From the divergences of $\vec{E}^a$ and $\vec{B}^a$,
it is possible to conclude that there exist sources of electric
and magnetic color fields which are not fermions. It is possible
to see that the bosonic fields are charged and simultaneously are
sources of magnetic fields, i. e. the gluonic fields have color
charge. Moreover, as the divergence of $\vec{B}^a$ non vanishing
then there exist color magnetic monopoles and the sources are the
gluons but not the quarks. As happens with the gauge potential in
electrodynamics, we have also found that the electric and magnetic
color fields are not gauge invariant. For these reason these
fields do not have any physical meaning.
\par
Finally, we have presented the integral formulation of non-abelian
Maxwell's equations by using both the Gauss-Ostrogradski theorem
and the Stokes theorem. For these, we have first built the
continuity equation and then we have introduced the
energy-momentum tensor as a function of electric and magnetic
color fields. We have found that the energy-momentum tensor has
the same interpretation as in electrodynamics.

\section*{Acknowledgments} This work was supported by COLCIENCIAS
(Colombia) under the research grant 1101-05-13610, CT215-2003.

\subsection*{Appendix : Wave equations}
Using the SU(3)-Yang-Mills equations Eq. (\ref{yms}) and the
definition of the non-Abelian gauge field tensor Eq. (\ref{ym1}),
it is possible to obtain
\begin{eqnarray}
\partial ^\mu F^ a _{\mu \nu}&=&-gC^a _{bc}A^{b \mu} F^c _{\mu \nu},\notag\\
\partial ^\mu(\partial _\mu A^a _\nu -\partial _\nu A^a_\mu +gC^a_{bc}
A^b_{\mu} A^c_\nu)&=&-gC^a _{bc}A^{b \mu} (\partial _\mu A^c _\nu -
\partial _\nu A^c_\mu +gC^c_{mn}A^m_\mu A^n_\nu),\notag\\
\Delta A^a _\nu -\partial _\nu \partial ^\mu A^a_\mu +
gC^a_{bc}\partial ^\mu(A^b_{\mu})A^c_\nu&=&-gC^a _{bc}A^{b\mu}
(2\partial _\mu A^c _\nu -\partial _\nu A^c_\mu +gC^c_{mn}A^m_\mu
A^n_\nu).\hspace{1cm}
\end{eqnarray}
Using the Lorentz gauge (fixing the gauge), we obtain a wave
equation given by
\begin{eqnarray}
\Delta A^a _\nu&=&gC^a _{bc}A^b _\mu (\partial _\nu
A^c_\mu-2\partial _\mu A^c _\nu-gC^c_{mn}A^m_\mu A^n_\nu).
\end{eqnarray}

\end{document}